\documentclass[fleqn,usenatbib]{mnras}

\usepackage{newtxtext,newtxmath}

\usepackage[T1]{fontenc}
\usepackage{ae,aecompl}

\usepackage{graphicx}	
\usepackage{amsmath}	
\usepackage{amssymb}	

\title[Brown dwarfs in NGC2264]{The brown dwarf population in the star forming region NGC2264}

\author[S. Pearson et al.]{
Samuel Pearson$^{1}$,\thanks{E-mail: sp246@st-andrews.ac.uk}
Aleks Scholz$^{1}$,
Paula S Teixeira$^{1}$,
Koraljka Mu\v{z}i\'{c}$^{2}$,
\newauthor
Jochen Eisl{\"o}ffel$^{3}$
\\
$^{1}$ SUPA, School of Physics \& Astronomy, University of St Andrews, North Haugh, St Andrews, KY16 9SS, United Kingdom \\
$^{2}$ CENTRA, Faculdade de Ci\^{e}ncias, Universidade de Lisboa, Ed. C8, Campo Grande, 1749-016 Lisboa, Portugal\\
$^{3}$  Th{\"u}ringer Landessternwarte, Sternwarte 5, D-07778 Tautenburg, Germany
}

\date{Accepted XXX. Received YYY; in original form ZZZ}

\pubyear{2020}

\begin{document}
\label{firstpage}
\pagerange{\pageref{firstpage}--\pageref{lastpage}}
\maketitle

\begin{abstract}
The brown dwarf population in the canonical star forming region NGC2264 is so far poorly explored. We present a deep, multi-wavelength, multi-epoch survey of the star forming cluster NGC2264, aimed to identify young brown dwarf candidates in this region. Using criteria including optical/near-infrared colours, variability, {\it Spitzer} mid-infrared colour excess, extinction, and {\it Gaia} parallax and proper motion (in order of relevance), we select 902 faint red sources with indicators of youth. Within this sample we identify 429 brown dwarf candidates based on their infrared colours. The brown dwarf candidates are estimated to span a mass range from 0.01 to 0.08\,$M_{\odot}$. We find rotation periods for 44 sources, 15 of which are brown dwarf candidates, ranging from 3.6 hours to 6.5 days. A subset of 38 brown dwarf candidates show high level irregular variability indicative of ongoing disc accretion, similar to the behaviour of young stars.


\end{abstract}




\section{Introduction}

Brown dwarfs are substellar objects intermediate between stars and planets, with masses too low to sustain stable hydrogen fusion ($M<0.08\,M_{\odot}$). They are a natural outcome of the processes that lead to the formation of stars and planets, and we find them in large numbers in every environment that has been probed so far with sufficient depth, star forming regions \citep{luhman2006survey}, open clusters \citep{openC}, the solar neighbourhood \citep{burningham201047}, the Galactic halo \citep{halo}, as companions to stars, and probably soon in globular clusters as well \citep{Dieball}. Overall, about 0.2-0.5 free-floating brown dwarfs exist per star, i.e. tens of billions across the Milky Way \citep{muzic}, with masses ranging from the substellar limit down to a few Jupiter masses.

The origin of brown dwarfs has long been a subject for debate. Given that stars and planets form in very different ways, the formation of brown dwarfs was initially discussed on the spectrum between `star-like' and `planet-like', i.e. either from the collapse of clouds \citep{bate} or within discs, followed by an ejection \citep{stamatellos}. Today, it is consensus view that most of the brown dwarfs with relatively high masses ($>0.02\,M_{\odot}$) form `like stars' \citep{luhman2012}, but the exact mechanism that stops them from growing further and the processes that determine their numbers and the shape of the very low mass IMF, remain unclear \citep{muzic2019}.

The prerequisite for making progress on this issue is to have large and well characterised samples of young brown dwarfs, in diverse star forming regions. A number of groups have conducted deep brown dwarf surveys of the nearest regions (within 350\,pc) over the past decade or so \citep{Scholz2012,muzic2013sonyc,Luhman_2000,Oliveira,pena2012new}. While these regions have obvious advantages for detailed characterisation, they do not provide the numbers, with typically less than 100 substellar objects (the exception being the dispersed, and slightly older, Upper Scorpius region \citep{Lodieu2}. We have recently extended this work to more distant and more extreme regions \citep{muzic2017,muzic2019}, but spectroscopic follow-up is a challenging proposition for young brown dwarfs beyond 1\,kpc. One well-studied cluster that has not been looked at thus far in sufficient depth to identify brown dwarfs down to planetary masses, is NGC2264 \citep{2264_bd1, 2264_bd2, 2008PhDPaula, extinctionMap}. With more than 1500 young stars known, we expect this cluster to host hundreds of substellar objects, and with a distance of 719 $\pm$ 16 pc the region is still accessible for detailed follow-up \citep{apellaniz2019gaia}.

In this paper, we set out to systematically explore the brown dwarf population of  NGC2264. The study is based on deep, multi-epoch imaging with the Blanco 4-m telescope on Cerro Tololo, which we use to create an initial object catalogue and lightcurves, complemented by near-IR photometry from the Florida Multi-object Imaging Near-IR Grism Observational Spectrometer (FLAMINGOS) \citep{elston, levine} (Section \ref{blanco}). This catalogue was enhanced with data products from Gaia DR2 and The \emph{Spitzer} Space Telescope (Section \ref{archive}). We identify brown dwarf candidates with this multi-faceted dataset, using optical/near-infrared colours, variability, infrared excess, extinction, and in some cases proper motions as selection criteria (Section \ref{selection}). The resulting sample, is presented and discussed in Section \ref{discussion}.

\section{The optical/near-infrared catalogue}
\label{blanco}

\subsection{Blanco observations}

The basis of this investigation are observations obtained with the Victor Blanco 4-m telescope at the Cerro Tololo Interamerica Observatory (CTIO) in Chile. We observed over six nights from 18/01/2010 to 23/01/2010, as part of NOAO proposal 2009B-0090, using the 64 Megapixel MOSAIC II camera. This instrument is comprised of eight CCDs, each with 2048x2048 pixels, in a $2\times4$ array. It has a field of view of $36'\times 36'$, corresponding to a pixel scale of 0.27\,$"$/pixel. The field was selected to cover the majority of NGC2264, centered on $(\alpha, \delta) (J2000) = (06^h42^m03.9^s, +09^{\circ}53'30.9")$. The field of view is large enough to encompasses the entire cluster from the cone nebula in the south-west to S Mon in the north-east. The pointing was the same for all images, within $5"$. Only seven of the detector's eight CCDs were operational during the period of the observations; the broken CCD in the northwestern corner of the array was used to block out S\,Mon, a bright O7-B-B triple system \citep{mon15} that would have otherwise saturated a large area of the detector. We observed in the Mosaic I-band filter (c6028), which has a central wavelength of 805\,nm and a FWHM of 150\,nm. 
A summary of the observations is given in table \ref{tab: Observations}. An average of 37 images were captured per night, each with an exposure time of 300\,s, resulting in a total of 221 images over the six night observing run.

\begin{table}
\centering
\begin{tabular}{|c|c|c|c|c|c|}
\hline
Date       & $t_{\mathrm{exp}}$ (s)    & no. & $\Delta t$ (h) & seeing (arcsec) \\ \hline
18/02/2010 & 300 & 32  & 04:13 & 0.8 - 34 \\ 
19/02/2010 & 300 & 37  & 04:55 & 0.9 - 11.9 \\ 
20/02/2010 & 300 & 36  & 05:03 & 0.8 - 10.6 \\ 
21/02/2010 & 300 & 42  & 05:04 & 0.9 - 1.3 \\ 
22/02/2010 & 300 & 34  & 05:18 & 0.8 - 1.2 \\ 
23/02/2010 & 300 & 40  & 05:13 & 0.7 - 1.3 \\ \hline
\end{tabular}
\caption{A summary of the I-band observations made with BLANCO telescope, including observing date, exposure time, number of images obtained, time covered, and range of seeing.}
\label{tab: Observations}
\end{table}

\subsection{Blanco stacking and source detection}

To create a deep stacked image and lightcurves, we made use of the single-frame \textit{Resampled} images, which we retrieved from the NOAO Science Archive (\textit{http://archive.noao.edu}) on 19/02/2019. \textit{Resampled} images are classed as level-2 products by the NOAO processing pipeline. In these images, the instrumental signature has been removed and geometric and photometric calibrations have been applied. This includes artifact flagging, cross-talk correction, bias correction, fringe pattern correction and flat-fielding. The processing pipeline also reprojected each image to a common grid using a sinc interpolation, and applied a world coordinate system by associating the centroids of stars in the science images with objects in an astrometric catalog. Details of the reduction and correction steps can be found in the CTIO MOSAIC II imager user manual \citep{mosaic}. 

\begin{figure}
	\centering
  	\includegraphics[width=0.45\textwidth]{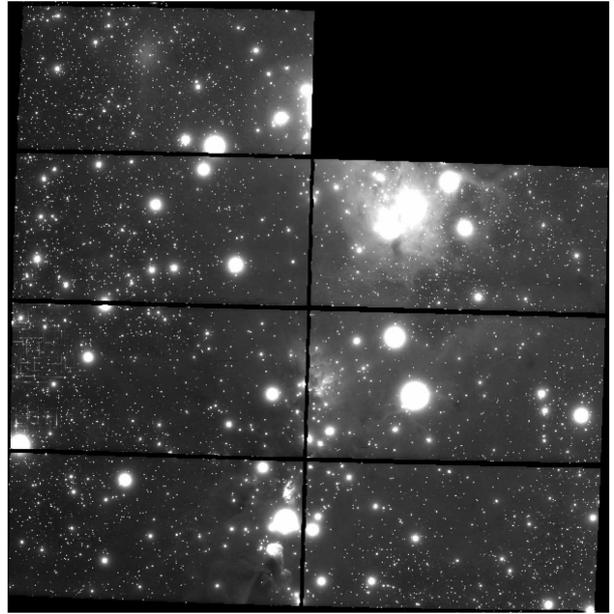}
	\caption{The resulting stacked image of 214 I-band exposures of NGC2264.}
	\label{fig:N2264}
\end{figure}

The task \textit{Imcombine} in IRAF \citep{iraf} was used to align and stack 214 of the 221 individual images. Seven of the images were rejected as bad images with greater than 2$"$ seeing. It was found that the WCS applied by the NOAO pipeline was accurate to within less than pixel (0.27$"$), sufficient to align the images during the stacking process. The resulting combined image (Figure \ref{fig:N2264}) is, to our knowledge, the deepest optical image of this region ever taken.

Next, we used \textit{DAOStarFinder} from photutils \citep{phot, astropy:2013} to detect sources in the deep I-band image. \textit{DAOStarFinder} searches the image for local density maxima that have a peak amplitude greater than a defined threshold, and have a size and shape similar to a 2D Gaussian kernel. A FWHM of 6 pixels was used for the kernel as this was found to be the typical FWHM of stars in the stacked image. \textit{DAOStarFinder} then locates the object centroid by fitting the x and y 1D distributions of the Gaussian kernel to the x and y distributions of the input image.

\textit{DAOStarFinder} searches for peaks exceeding a chosen threshold above the average background count rate for the whole image. Because the background is highly variable across the field, due to the nebula, we chose a very low cut-off of 0.1 standard deviations above the average background count rate, to ensure that sources lying in areas with higher than average background counts were not ignored. The catalogue based on the I-band image alone, thus contains a significant number of contaminants. In total, the I-band catalogue contains 85000 sources, intentionally designed to include all sources in the field (rather than being a particularly clean sample).

\subsection{Cross-referencing with FLAMINGOS}

The I-band source catalogue was then cross referenced with a catalogue of deep near-infrared J (1.25 $\mu$m), H (1.65 $\mu$m) $\&$ K (2.2 $\mu$m) photometry from the FLAMINGOS giant molecular cloud survey \citep{elston, levine}. The FLAMINGOS catalog  was chosen as it provided the deepest available near-infrared photometery for the cluster, the approximate completeness limits in J, H, and K are 19.2, 18.8, and 17.9 mag, see Figure \ref{fig:FullComp}. The saturation limits for FLAMINGOS derived from a comparison with 2MASS are 11.0, 11.5, and 12.0 mag for J, H, and K, respectively. Bright sources will typically become saturated in the MOSAIC-II I-band data well before The FLAMINGOS JHK data. The spatial coverage of the FLAMINGOS data is very similar to the MOSAIC-II data, with $>95\%$ overlap.

Any sources detected in the I-band image that did not have a match within 1$"$ of the coordinates of an object in the FLAMINGOS catalogue were discarded. This serves two purposes, on one hand it effectively removes spurious sources from the I-band catalogue, on the other, it only retains sources with multi-band information, which we need for characterisation. One drawback to this method was that any sources that were not in the FLAMINGOS catalogue will be missed, and will not be included in our catalogue. This also included sources that do not lie within the overlap of the two surveys, as they had slightly different fields of view, as well as objects that were too faint to be detected in JHK. The completeness of our survey will be determined by the combination of these two catalogues. In total, 18,196 objects were identified as matches between the Blanco and FLAMINGOS catalogues. 



\subsection{Blanco photometry}

For the sources in the combined Blanco/FLAMINGO catalogue we carried out aperture photometry on the stacked I-band image, using again the photutils package \citep{phot, astropy:2013}. An aperture of 5 pixel radius (1.35$"$) was positioned on the centroid position to measure the flux from each object. The local median background count was calculated and subtracted for each source by measuring the median in an aperture ring (outer radius 15 pixels, inner radius 10 pixels). 

\begin{figure}
	\centering
  	\includegraphics[width=0.45\textwidth]{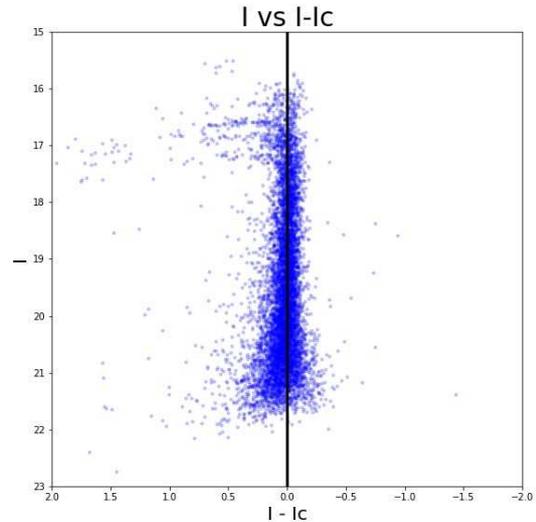}
	\caption{A comparison between the Lamm catalogue and our calibrated Blanco} photometry showing they are in good agreement. The structure in the top left side of the figure is an artifact from the saturated sources in the Blanco dataset. Here I is the Blanco I-band filter and Ic is the cousins I-band filter used by \citet{Lamm}
	\label{fig:LammComp}
\end{figure}

For calibration purposes, the magnitudes of our sample were then compared to the I-band photometry from \citet{Lamm} survey of the same region. A total of 7702 Lamm catalogue objects were found to match one of our sources within 1$"$ (see Figure \ref{fig:LammComp}). The average difference between our magnitudes and those measured by Lamm was subtracted from our I-band magnitudes. This means that the accuracy of the calculated apparent magnitudes is reliant on the accuracy of the Lamm catalogue magnitudes. We note that Lamm's photometry was calibrated against standard stars from \citet{rebull02}.

A significant problem we encountered during the photometry was saturation. We identified and removed 2354 saturated (and thus non-circular) sources using the `roundness' parameters in photutils. We call the remaining 15,735 sources the BLANCO sample. From comparison with the Lamm catalogue, in addition to examining the images themselves and the data quality masks that accompany the data from the NOAO Science Archive, it was determined that any sources brighter than 17th magnitude in the BLANCO sample suffered from some degree of saturation. These sources were removed from the catalogue and replaced with the Lamm catalogue entries. Other Lamm sources brighter than the range of our catalogue ($<12$ mag), but with matching FLAMINGOS JHK magnitudes, were also added for completeness. The final catalogue has 16,317 sources. We call this sample the BASE sample.

\begin{figure}
	\centering
  	\includegraphics[width=0.45\textwidth]{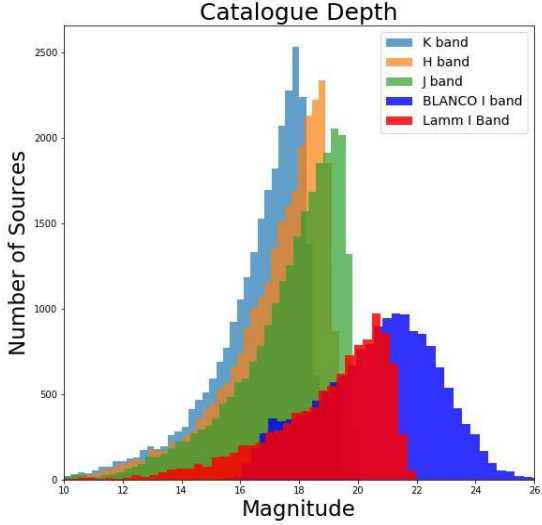}
	\caption{A histogram of the I-band magnitudes of the BLANCO and the Lamm catalogue, as well as the JHK magnitudes in the BASE catalogue.}
	\label{fig:FullComp}
\end{figure}

In Figure \ref{fig:FullComp}, we plot histograms of the I-band magnitude distribution for our sample and the Lamm catalogue. As apparent from this figure, our catalogue is substantially deeper than previous work in this cluster, with a limiting magnitude below 25 and a completeness limit (defined as the peak of the histogram) around 22 mag. For comparison, the histogram of the Lamm catalogue peaks around 20.5 mag. In terms of masses of substellar members of the cluster, this means that our catalogue has the potential to include objects down to planetary masses, whereas previous work does not extend as far into the substellar domain. In Figure \ref{fig:FullComp} we also show the histogram of the near-infrared photometry from the FLAMINGOS catalogue.

\subsection{Time series photometry}

The presence of a disc, as well as ongoing accretion and magnetic activity can induce optical variability, in particular periodic modulations due to spots or warps in the disc. Accretion can also cause irregular and high amplitude variability. Previous work has shown that this type of variability is indeed observed in young brown dwarfs \citep{aleksVar,aleksVar2}. Periodic or high-level variability among candidates can therefore be used as additional confirmation of youth, and thus cluster membership.

The BLANCO sample was used to conduct multi-epoch photometry on each of the 221 individual images. We used the WCS associated with each individual image to convert the RA and DEC for each source in the catalogue to a pixel coordinate that could then be used to run aperture photometry again using photutils. As previously mentioned, the WCS was found to be accurate to within a pixel (0.27"), i.e. no re-centering from one image to the next was needed. Sources fainter than 24 mag were not robustly detected in the individual images and were discarded. Sources fainter than 22 mag were not used for any part of the lightcurve calibration.

We carried out differential calibration following the method detailed in \citet{rot}. This method selects a set of non-variable stars and then subtracts the average lightcurve of the set from those of all sources. This corrects for variation due to fluctuating atmospheric conditions, changing airmass, or changing instrumental sensitivity. 

We include here a short summary of the method, for more details see \citet{rot}. The source number is indicated with $i = 1...N_{S}$, the image number with $j = 1...N_{I}$.


\textsc{1) Identifying bad images:}

\begin{enumerate}
    \item For each source, the average magnitude over all 221 images was calculated.
    
    \begin{equation}
        \overline{m}_i = \frac{1}{N_{I}}\sum^{N_{I}}_{j=1}m_i(t_j)
    \end{equation}
    
    where $t_j$ is epoch $t$ of image $j$
    
    \item The average magnitude was then subtracted from every value for the time series, for every individual source.
    
    \begin{equation}
        m^0_i (t_j)  = m_i(t_j) - \overline{m_i}
    \end{equation}
    
    \item For each image the average and standard deviation of the average subtracted values were found.
    
    \begin{equation}
        \overline{m}^0_{j} = \frac{1}{N_{S}}\sum^{N_{S}}_{i=1}m^0_i(t_j)
    \end{equation}
    
    \begin{equation}
        \sigma_j = \sqrt{\frac{1}{N_S - 1} \sum^{N_S}_{i=1}(m^0_i (t_j) - \overline{m^0_{j})}^2}
    \end{equation}

    \item 9 `bad' images with high standard deviations ($\sigma_j > 0.5$) were identified and excluded from the following.
\end{enumerate}

\textsc{2) Selecting the non-variable stars:}

The next task was to select a pool of non-variable reference stars. This was achieved by calculating the quality test number $test_i$. For every source in each image, the quality test number compares the variation of a source from its average magnitude, to that of every other source, taking into account the overall average variation of each image. 

\begin{enumerate}
    \setcounter{enumi}{4}
    \item Steps (i)-(iii) were repeated with the 9 \textit{bad} images removed.
    
    \item The quality test number $test_{ij}$ was calculated.
    
    \begin{equation}
        \textsc{if} \;\;  |m^0_i (t_j) - \overline{m^0_{j}}| \geq \sigma_j, \;\;\; \textsc{then} \;\; test_{i,j} = 1, \;\;\; \textsc{else} \;\; test_{i,j} = 0
    \end{equation}
    
    \item The sum of these quality test number was calculated.
    
    \begin{equation}
        test_i = \sum_{j=1}^{N^{'}_{I}} test_{i,j}
    \end{equation}
    
    183 sources with $test_{i} = 0$ were selected as non-variable reference sources.
   
\end{enumerate}

\textsc{3) Calibrating the lightcurves:}

The final step was to use the bright, non-variable reference stars to calculate the mean lightcurve and subtract it from each source.

\begin{enumerate}
    \setcounter{enumi}{7}
    \item For each image, the average brightness of the reference stars was calculated in order to find the `mean lightcurve'.

    \begin{equation}
        \overline{m^{ref}}(t_j) = \frac{1}{N_{ref}}\sum_{i=1}^{N_{ref}}m^{ref}_{i}(t_j)
    \end{equation}

    \item The mean lightcurve was then subtracted from every source, resulting in calibrated lightcurves.
    
    \begin{equation}
        m^{rel}(t_j) = m(t_j) - \overline{m^{ref}}(t_j)
    \end{equation}

\end{enumerate}

\begin{figure}
	\centering
  	\includegraphics[width=0.45\textwidth]{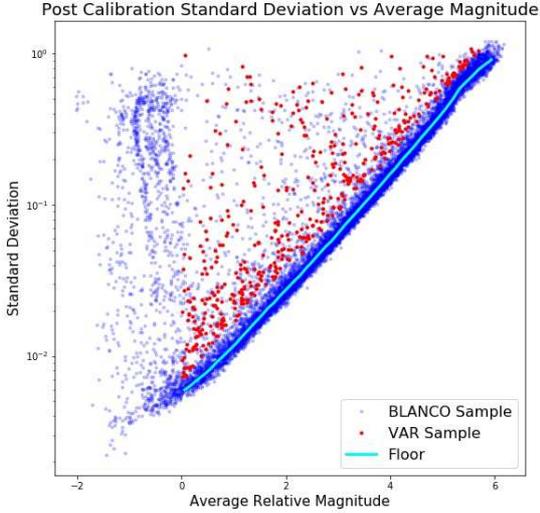}
	\caption{A plot comparing the average magnitude and standard deviation of each source over the 221 images, after the differential photometry and calibration.}
	\label{fig:LC1}
\end{figure}

This finishes the treatment of the lightcurve. Figure \ref{fig:LC1} shows the standard deviation vs average magnitude of the sources after this process. The average and standard deviation for each source in this diagram was calculated using an outlier rejection with 3$\sigma$ as threshold. The `floor' in this diagram describes the photometric noise as a function of relative magnitude. The photometric noise reaches a minimum of 0.003 mag for bright stars. At the substellar limit, corresponding to approximately 0.5 in relative magnitude, it reaches 0.006 mag. At the bright end, the diagram shows obvious structures. We examined this in depth and found that these structures are caused by saturated sources in the time series catalogue. None of these sources were used in the calibration, and none are faint enough to be considered in our candidate selection.


\section{Complementary archival datasets}
\label{archive}

\subsection{Gaia DR2}
We make use of the second data release Gaia DR2, Gaia Collaboration \citealt{gaia1, gaia2}. Gaia DR2 was found to contain proper motions and parallaxes for 4811 matching sources in our BASE catalogue. We also utilised the distance to NGC2264 of 719 $\pm$ 16 pc, determined by \citet{2264dist} using Gaia DR2. We note that while very useful for accurately determining the distance to NGC 2264, Gaia DR2 data is very limited for brown dwarf kinematics in this cluster as only the very brightest brown dwarfs $(I\sim17.5)$ are reached.

\subsection{Spitzer}


In 2004, NGC2264 was observed with the \emph{Spitzer} Space Telescope as part of Guaranteed Time Observation program 37. Here we make use of a catalogue of 6548 objects identified in this survey by \citet{extinctionMap}. This catalogue is comprised of the IRAC photometry in four bands, centred at 3.6$\mu m$, 4.5$\mu m$, 5.8$\mu m$ and 8$\mu m$. Of the 6548 objects, 2092 were found to match sources in our BASE catalogue within 3$"$. This is a useful sample for our survey as infrared excess can be used to identify sources with discs, which is a clear indication of youth. The \emph{Spitzer} photometry spans the full range of stars and brown dwarfs, down to $(I\sim25)$.

\section{Selection of candidate brown dwarfs in NGC2264}
\label{selection}

In this section we describe the selection of substellar members in NGC2264 using our multi-faceted dataset.

\subsection{Colour-Magnitude}

\begin{figure}
	\centering
  	\includegraphics[width=.45\textwidth]{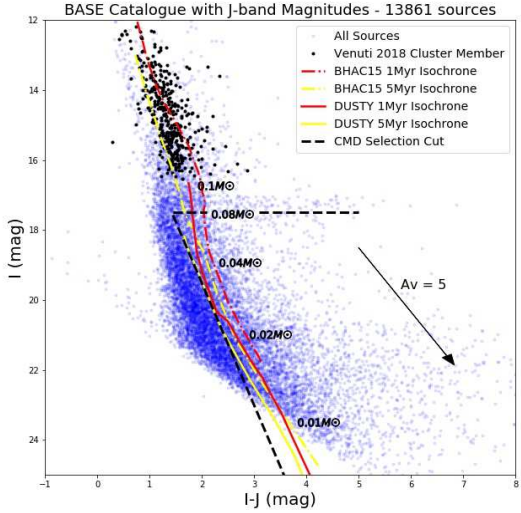}
	\caption{I vs I-J colour magnitude diagram for the BASE catalogue, overplotted with known cluster members and 1 \& 5 Myr BHAC15 \& DUSTY isochrones. An approximate mass scale is labelled in black. The straight line cutoff at the faint end of the catalogue is due to the limiting depth of the FLAMINGOS catalogue.}
	\label{fig:CombCatCMD}
\end{figure}

The first step in the selection of candidate substellar members is the examination of optical/near-infrared colours. An I vs I-J colour magnitude plot of our catalogue is shown in Figure \ref{fig:CombCatCMD}. The known stellar members from \citet{sequential} are shown in black. BHAC15 \citep{Baraffe} and DUSTY isochrones \citep{DUSTY} for 1 and 5Myrs have been overplotted using a distance of 719 pc to convert between absolute and apparent magnitude. Both 1 and 5Myrs are shown to account for the suspected age spread in NGC2264 \citep{sequential}. These isochrones were also used to apply an approximate mass scale. Note that the isochrone is unreddened, i.e. masses correspond to $A_V=0$. These isochrones reproduce the position of the stellar members in this diagram very well, our selection criterion will select objects in the colour-magnitude space at the faint extension of these known stellar sources. The young sources in NGC2264 are expected to lie on and to the right of the isochrones, due to reddening. Of the 13,861 sources with J-band magnitudes in the BASE catalogue, 6,200 of them lie in the expected region for young brown dwarfs, shown by the black dashed line in Fig \ref{fig:CombCatCMD}. We defined this region with two cuts, fainter than 17.5 magnitude in I, equivalent to approximately $0.08M_\odot$. To the right of the line $y = 3.5x + 12.5$ (a conservative approximation of the isochrones), equivalent to sources that are redder than the isochrone colours at $A_V=0$. We call this sample the CMD sample. The CMD sample will be the most complete sample, but contains a large proportion of contaminants from reddened background sources and embedded stars. The additional selection criteria discussed in the sections below are used to help distinguish young and very low mass sources.


\subsection{Variability}

Variable sources were identified using Figure \ref{fig:LC1}. For any given magnitude in our sample there is an inherent photometric noise, measured here as the standard deviation over the lightcurves, after differential calibration. The photometric noise, which includes photon noise, background noise, and systematics, increases for fainter sources and can be seen as the `floor' in Figure \ref{fig:LC1}. This allows for an easy way to select variable sources as they will exhibit a greater variation over the six nights and therefore be found above the `floor', defined by the bulk of the objects in the field.

We determined the `floor' by first selecting a subsample of sources likely to be constant, i.e the main bulk of sources forming the straight line in Figure \ref{fig:LC1}. This subset of data was then binned. The average of the photometric standard deviations in each bin was used to plot the `floor' in Figure \ref{fig:LC1}. All sources that were found to lie 5 standard deviations above the average for their bin were selected as possible variable sources. In total, 591 sources from the CMD sample were found to meet this criterion. 

Many of these sources show obvious variability on timescales of hours or days, when examining the lightcurves visually. However, some lightcurve seem to show merely an increased level of white noise. To find and reject those lightcurves, we used an autocorrelation test. Autocorrelation is the correlation of a signal with a delayed copy of itself. For a random signal, the autocorrelation should be close to zero for the majority of lags. We used four lags of 1-4 steps along a lightcurve to define the delayed signal, equivalent to shifting the lightcurve along by 1-4 data points. Sources where the autocorrelation values for all four lags were close to zero (between -0.05 \& 0.05), were rejected as white noise. In all, 104 sources were rejected as white noise. In total 467, sources were classified as young candidates based on their variability. This sample is in the following referred to as VAR sample.

Visual examination shows that a subsample of this group is highly variable with amplitudes of 0.2 - 2 magnitudes. This sample of 83 sources is in the following referred to as HIGHVAR. Sources with red colours and variability on the timescales covered here are likely to be young, as variability in background stars on timescales of days is not expected to be common. This is particularly valid for the HIGHVAR sample. 

\subsection{Gaia}

\begin{figure}
	\centering
  	\includegraphics[width=0.45\textwidth]{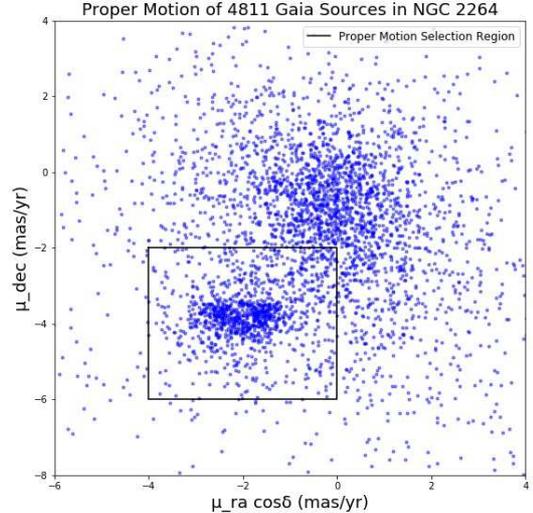}
	\caption{The proper motions of 4811 sources within NGC2264. The smaller cluster of sources is NGC2264, the black box indicates the proper motion selection criteria.}
	\label{fig:gaia1}
\end{figure}

We attempted to use Gaia DR2 to determine low mass candidates based on their kinematics using the selection criteria outlined below:

\begin{itemize}
    \item a parallax between 1.3 and 1.5 mas (719 $\pm$ 50 pc)
    \item a proper motion consistent with the locus of NGC2264 (see Figure \ref{fig:gaia1})
    \item parallax error less than 30\% 
\end{itemize}

In total, 24 sources were found to satisfy these criteria and were also in the CMD sample. This sample is in the following referred to as KINEMATIC sample. As the Gaia errors become increasingly large at these faint magnitudes, these 24 objects are all found around the expected threshold between stars and brown dwarfs. 

\subsection{Spitzer}

\begin{figure}
	\centering
    \includegraphics[width=.45\textwidth]{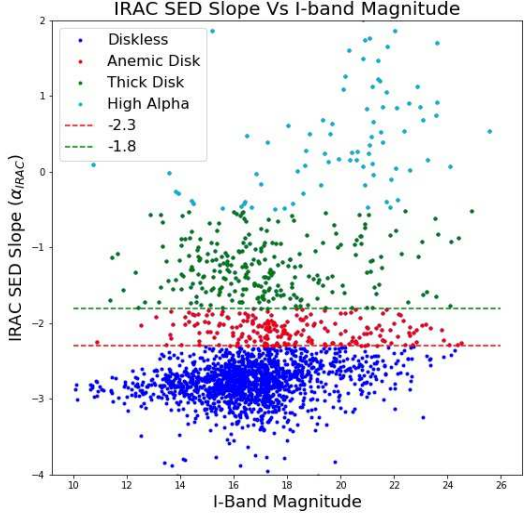}
	\caption{The slope of the Spitzer/IRAC SED $\alpha_{IRAC}$ vs. I-band magnitude. he cuts used to differentiate between disc bearing and discless sources are taken from \citet{Lada_2006}.}
	\label{fig:SpitCands}
\end{figure}

We expect that most members of NGC2264 will still have circumstellar discs due to their young ages. The presence of a disc can be inferred from infrared excess with respect to the stellar photosphere. The slope of the infrared spectral energy distribution (SED) was used to identify sources with an excess indicative of a disc and to differentiate between the sources with thick discs and anaemic discs \citep{Lada_2006} \citep{extinctionMap}, see Figure \ref{fig:SpitCands}. Sources with $-2.3 < \alpha_{IRAC} < -1.8$ were classified as having anaemic discs and sources with $-1.8 < \alpha_{IRAC} < -0.5$ as thick discs. Sources with very high alpha ($\alpha_{IRAC} > -0.5$) are likely to be highly embedded, they may be edge on discs, protostars or flat spectra. For this study, we have chosen not to include these sources as possible young candidates. In total, 199 sources show infrared excess and are also found in CMD, this sample is in the following referred to as DISCS. Of the identified sources, 111 have thick discs, whereas 88 have anaemic discs. This selection should reliably exclude most red background objects from the initial CMD sample. The DISCS sample is expected to be comprised of mostly young sources, but some could be embedded young very low mass stars with high extinction, rather than brown dwarfs. This criterion will also exclude young sources without discs, and hence will not provide a complete census.


\subsection{Extinction}

\begin{figure}
	\centering
  	\includegraphics[width=0.45\textwidth]{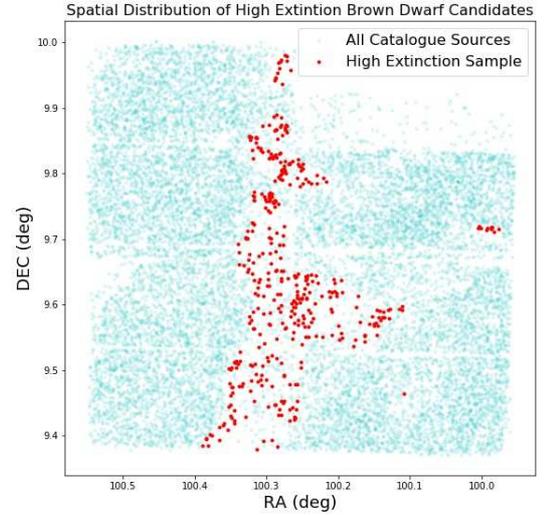}
	\caption{The spatial distribution of the 365 high extinction sample}
	\label{fig:HIGHEX}
\end{figure}

We also probed the spatial location of the sources and used that as an additional selection criteria. If $A_V$ is sufficiently large, the cloud acts as a screen and blocks out background sources. Thus, objects located in regions of very high extinction should be more robustly identified as members of the cluster. We made use of the NIR extinction map from \citet{extinctionMap}, built using HK Flamingos data and the near-infrared colour excess method \citep{alves1998dust}. For each source from CMD sample, the extinction level in the nearest 9 pixels of the extinction map was averaged, corresponding to a 75$"$ x 75$"$ area. Sources that lay within an area of $A_V > 10$ mean extinction level were flagged. In total, 365 of the CMD sources were flagged. This sample is in the following referred to as HIGHEX sample (Fig. \ref{fig:HIGHEX}).

\subsection{Infrared Colours}

\begin{figure*}
	\centering
  	\includegraphics[width=0.95\textwidth]{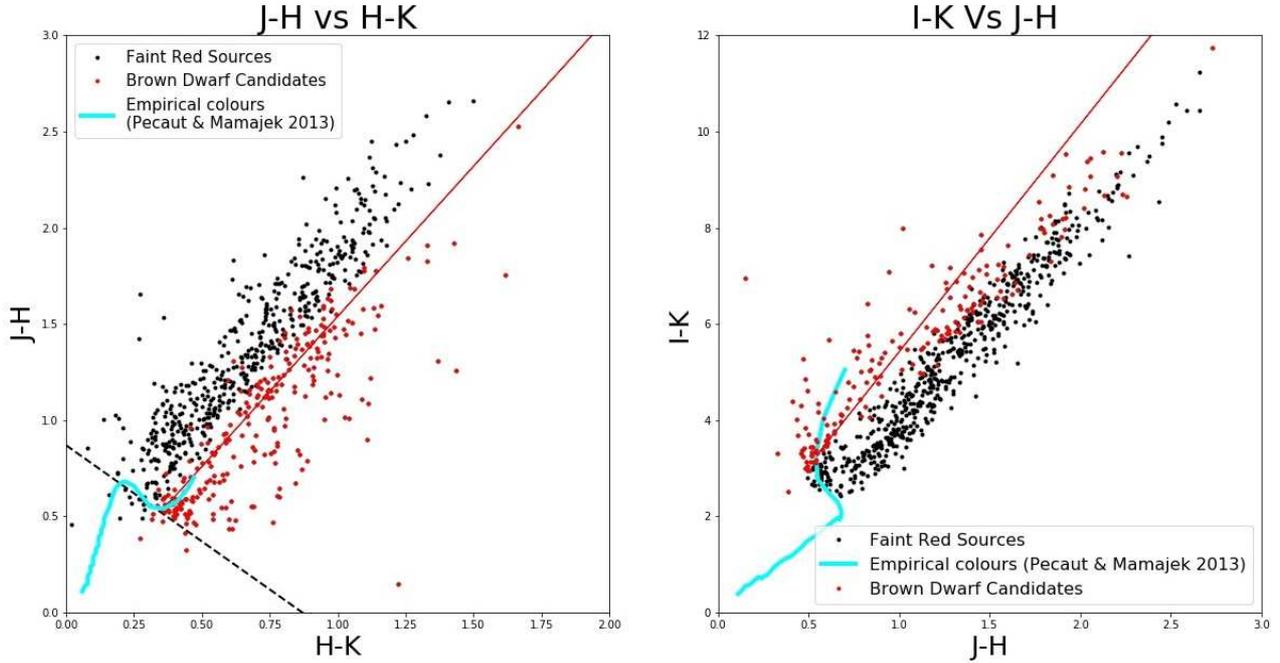}
	\caption{Two infrared colour-colour plots showing the 902 faint red sources identified in at least one of the samples: DISCS, KINEMATIC, HIGHEX \& VAR. The red lines indicate M6 extinction vectors. The extinction law was taken from \citet{extin} ($R_V = 3.1$), the empirical colours are from \citet{colourseq}. Sources falling in the expected regions for substeller objects, with their associated error bars, were selected as brown dwarf candidates. The dashed black line in the left plot is the approximation of the empirical colour sequence for young stars, used for extinction estimation.}
	\label{fig:ColColDub}
\end{figure*}

Two infrared colour-colour plots (J-H vs H-K \& I-K vs J-H) are shown in Fig \ref{fig:ColColDub}. The 902 sources shown in these plots are a combination of the samples DISCS, KINEMATIC, HIGHEX \& VAR and were therefore identified in at least one of the four methods in sections 4.2-4.5. From this sample of 902 faint red sources we identify brown dwarf candidates that are consistent with substellar colors ($>M6$). The red lines in these plots indicate M6 reddening vectors constructed using the extinction laws from \citet{extin} ($R_V = 3.1$) and the M6 empirical colours from \citet{colourseq}. Sources lying below the reddening vector in the JH/HK diagram and above the reddening vector in the IK/JH diagram have colours consistent with substellar objects and were selected as brown dwarf candidates. Sources that were found to fall in these regions within their associated error bars were selected as brown dwarf candidates. 273 sources were found to meet this condition in the JH/HK plot and 190 sources for the IK/JH plot.

\begin{figure}
	\centering
  	\includegraphics[width=0.45\textwidth]{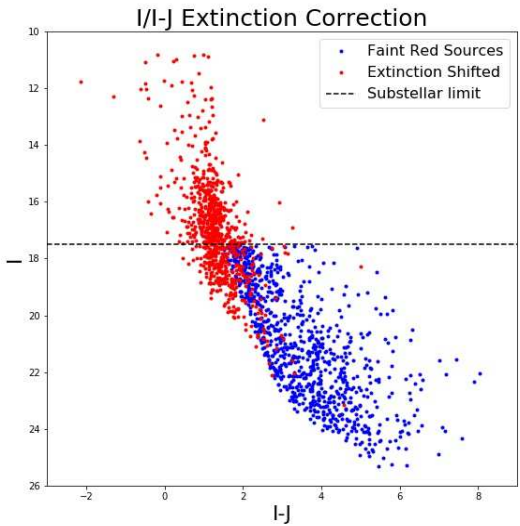}
	\caption{I/I-J colour-magnitude diagram for the 902 faint red sources identified in at least one of the samples: DISCS, KINEMATIC, HIGHEX \& VAR. Sources are shown both before and after an extinction correction. The extinction for each source was estimated using Fig \ref{fig:ColColDub}. Sources found to fall in the expected region for brown dwarfs post correction, were selected as brown dwarf candidates.}
	\label{fig:IJcorr}
\end{figure}

The JH/HK plot (Figure \ref{fig:ColColDub}) was also used to estimate extinction. The advantage of this colour-colour diagram is that the chosen colours do not depend strongly on mass. The extinction for each source was estimated by following the reddening vectors for $R_v=3.1$ \citep{extin} until they reached the dashed black line in Figure \ref{fig:ColColDub}. This line was used to approximate the locus of the empirical colour sequence for young stars \citep{colourseq} \& 3 Myr model isochrone \citep{Baraffe}. The estimated extinction was used to de-redden sources in the I/I-J colour-magnitude diagram (Fig \ref{fig:IJcorr}), sources that still fell in the expected region for substellar objects were then selected as brown dwarf candidates. 289 sources were selected.

In total, 429 sources were identified as brown dwarf candidates, with 105 of the candidates common to all three selections. We have kept sources from all plot as this is only a preliminary selection, we expect to find some brown dwarfs on the other side of our M6 reddening vector, and some stars in this subsample.



\subsection{Summary of selection criteria}

In this subsection we collate the candidate selection criteria to define our final sample of brown dwarf candidates. From the BASE catalogue we have selected: CMD, DISCS, KINEMATIC \& HIGHEX. From BLANCO we have selected VAR \& HIGHVAR. In total we select 902 faint red sources that are identified at least one of the four samples (DISCS, KINEMATIC, HIGHEX \& VAR). From these sources we have selected the 429 Brown Dwarf Candidates by their infrared colours. This is summarised in Table \ref{tab:Sum}. Sources that are found in several of samples can be treated with more confidence, as there are multiple lines of evidence pointing to their cluster membership, youth and substellar nature.


\begin{table*}
 \caption{Catalogue summary table}
 \label{tab:Sum}
 \begin{tabular}{lll}
  \hline
  Catalogue Name & Source Count & Progenitor\\
  \hline
  BASE & 16317 & \\[2pt]
  BLANCO & 15735 & \\[2pt]
  CMD & 6200 & BASE\\[2pt]
  DISCS & 199 & CMD\\[2pt]
  KINEMATIC & 24 & CMD\\[2pt]
  HIGHEX & 365 & CMD\\[2pt]
  VAR & 455 & BLANCO, CMD\\[2pt]
  HIGHVAR & 84 & VAR\\[2pt]
  ROT & 44 & DISCS, VAR\\[2pt]
  \hline
  JHK\_Sel & 273 & DISCS, KINEMATIC, HIGHEX, VAR\\[2pt]
  IKJH\_Sel & 190 & DISCS, KINEMATIC,HIGHEX, VAR\\[2pt]
  IJ\_Sel & 289 & DISCS, KINEMATIC, HIGHEX, VAR\\[2pt]
  Brown Dwarf Candidates & 429 & JKH\_Sel, IKJH\_Sel, IJ\_Sel\\[2pt]
  \hline
 \end{tabular}
\end{table*}




\section{The substellar population in NGC2264}
\label{discussion}

\subsection{Spatial distribution}

\begin{figure*}
	\centering
  	\includegraphics[width=.95\textwidth]{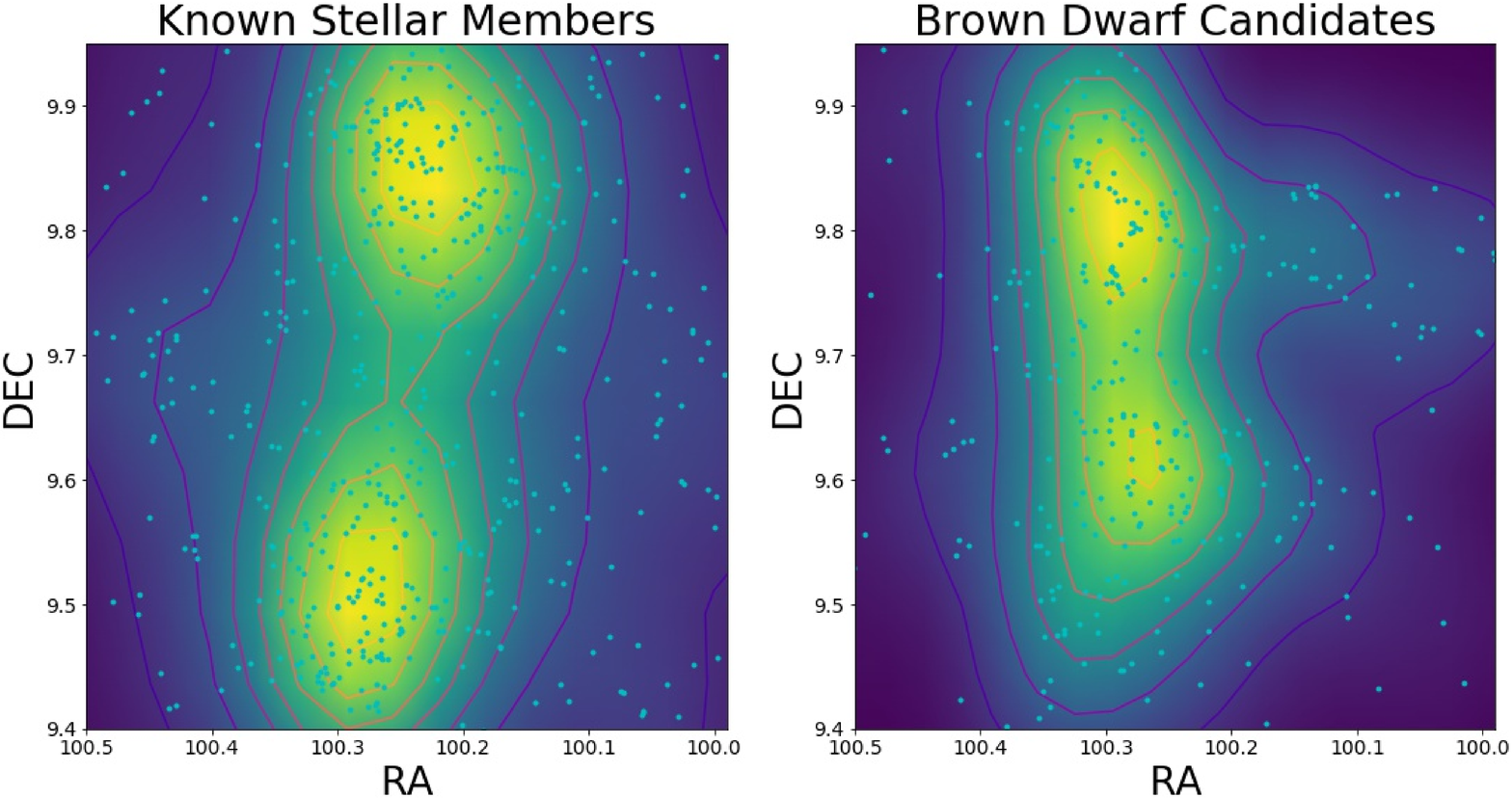}
	\caption{The Spatial Distribution of the Venuti et al. cluster members and brown dwarf candidates. The colour map and contours show the 2D density distribution calculated using a Gaussian kernel density estimate. The absence brown dwarf candidates in the upper right hand corner corresponds to the broken CCD chip from the BLANCO observations.}
	\label{fig:ProbMemSpat}
\end{figure*}

Figure \ref{fig:ProbMemSpat} shows the spatial distribution of 655 spectroscopically confirmed cluster members \citep{sequential} ranging in mass from 0.2 to 1.8\,$M_{\odot}$ and the 320 brown dwarf candidates. The colour map and contours in Figure \ref{fig:ProbMemSpat} show the 2D density distribution calculated using a Gaussian kernel density estimate. It can be seen from the known members that there are two main sub-clusters of active star formation within NGC2264, one to the north, close to the massive triple system S-Mon, and one to the south, near the tip of the cone nebula \citep{Paula2006, extinctionMap, sung2009, sequential}. We present here the spatial distribution of our brown dwarf candidates, but leave further analysis until candidates have been confirmed by spectroscopy.


\subsection{Rotation}
\begin{figure*}
	\centering
  	\includegraphics[width=.9\textwidth]{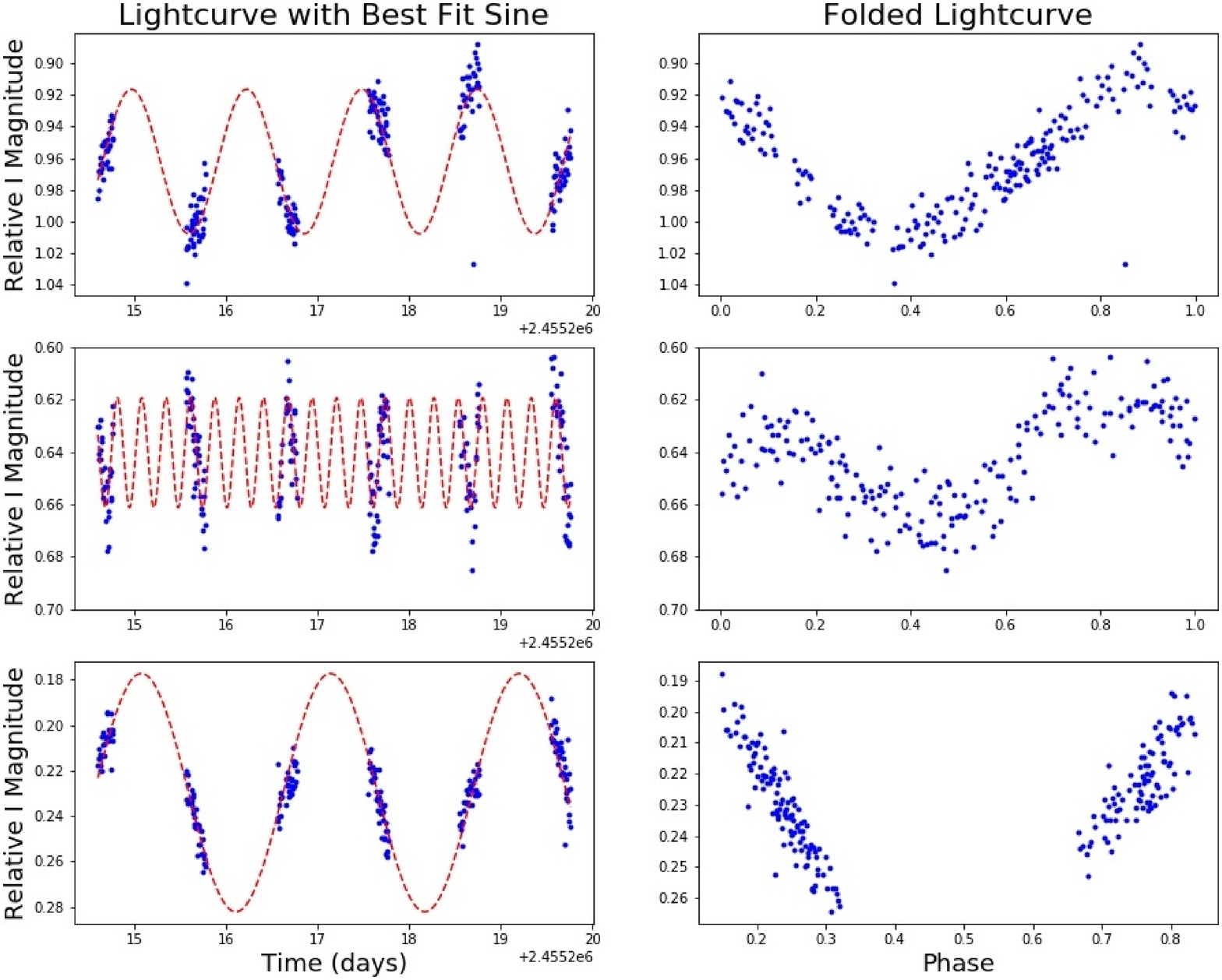}
	\caption{Example lightcurve and phase plots for sources fitted with Sine curves to determine the rotation period. In descending order: ID: 29, 23, 13 (Table \ref{tab:ROT}). Fitted rotation period (days): 1.26, 0.27, 2.06. Amplitude (magnitudes): 0.046, 0.021, 0.052.}
	\label{fig:Rot}
\end{figure*}

The lightcurves for sources in both the DISCS and VAR samples were examined using a generalized Lomb-Scargle periodogram, a common tool in the frequency analysis of unequally spaced data, equivalent to least-squares fitting of sine waves \citep{Zechmeister_2009}. We used the GLS class from PyAstronomy \citep{pya} to identify sources with very clear periodic variability, and fitted them with a simple sine curve to determine an approximate rotation period. Alongside the generalized Lomb-Scargle periodogram the GLS class incorporates the photometric error to calculate a False-Alarm-Probability (FAP). The photometric error for each source was calculated using its average magnitude and the equation for the `floor' of the photometric noise in section 4.3. All sources with a FAP less than 1\% were then examined by eye to determine whether the fit was appropriate for a periodic rotation.  A sinusoidal period detected in our lightcurve is almost certainly the rotation period \citep{Lamm}. In total, the rotation period of 44 sources was estimated, these are shown in Table \ref{tab:ROT}, 15 of these sources were selected as brown dwarf candidates. Examples of lightcurves overplotted with the best fit sine curve used to determine the period and the resulting phase plots, are shown in Figure \ref{fig:Rot}.

Of our sample of 44 sources with rotation periods, 19 of these were also identified as variable sources by \citet{Lamm}, 17 as periodic variable and 2 as irregular variables. Of the 17 periodic variables with \citet{Lamm} rotation periods, 12 of these match or are integer multiples of our periods, within an error margin of 10\%.

\begin{table}
 \caption{The rotation periods for the 44 ROT sources. The periods found by \citet{Lamm} are also shown, where iv refers to irregular varaible}
 \label{tab:ROT}
 \begin{tabular}{llllll}
 \hline
 Id & RA & DEC & I & $P$ (days) & $P_{Lamm}$ (days) \\
 \hline
 1  & 100.006 & 9.62903 & 17.58 & 2.15   & 1.81        \\
 2  & 99.9713 & 9.82422 & 17.58 & 6.44   & -           \\
 3  & 100.329 & 9.70228 & 17.59 & 0.63   & 0.63        \\
 4  & 100.017 & 9.78494 & 17.61 & 1.26   & 1.29        \\
 5  & 100.248 & 9.61603 & 17.62 & 6.44   & 5.43        \\
 6  & 100.299 & 9.50744 & 17.64 & 1.66   & -           \\
 7  & 100.136 & 9.58103 & 17.66 & 4.30   & 4.17        \\
 8  & 100.192 & 9.78734 & 17.70 & 1.20   & 2.35        \\
 9  & 99.9962 & 9.59792 & 17.72 & 1.47   & -           \\
 10 & 100.193 & 9.69910 & 17.73 & 0.64   & 1.9         \\
 11 & 100.046 & 9.75735 & 17.73 & 2.15   & 2.31        \\
 12 & 100.433 & 9.76446 & 17.73 & 2.24   & 2.22        \\
 13 & 100.122 & 9.75531 & 17.79 & 2.06   & -           \\
 14 & 100.196 & 9.50551 & 17.80 & 2.58   & 1.59        \\
 15 & 100.333 & 9.59335 & 17.86 & 1.43   & -           \\
 16 & 99.9696 & 9.70249 & 17.87 & 0.54   & -           \\
 17 & 100.293 & 9.76508 & 17.89 & 1.32   & -           \\
 18 & 100.425 & 9.95237 & 17.90 & 1.39   & -           \\
 19 & 100.170 & 9.72026 & 17.94 & 0.66   & -           \\
 20 & 100.016 & 9.75811 & 18.10 & 1.20   & -           \\
 21 & 100.327 & 9.81733 & 18.16 & 1.66   & -           \\
 22 & 100.396 & 9.41910 & 18.22 & 0.36   & 0.36        \\
 23 & 100.402 & 9.65767 & 18.22 & 0.27   & -           \\
 24 & 100.139 & 9.45791 & 18.24 & 1.17   & 1.04        \\
 25 & 100.330 & 9.89695 & 18.34 & 0.17   & 0.17        \\
 26 & 100.132 & 9.72517 & 18.38 & 0.59   & -           \\
 27 & 100.362 & 9.76678 & 18.47 & 3.03   & -           \\
 28 & 100.233 & 9.50482 & 18.51 & 0.73   & -           \\
 29 & 100.327 & 9.90213 & 18.53 & 1.26   & 1.24        \\
 30 & 100.205 & 9.45320 & 18.56 & 0.61   & -           \\
 31 & 100.183 & 9.76594 & 18.57 & 0.33   & -           \\
 32 & 100.058 & 9.56926 & 18.63 & 0.57   & -           \\
 33 & 100.418 & 9.80481 & 18.64 & 3.68   & -           \\
 34 & 100.347 & 9.54176 & 18.72 & 3.97   & -           \\
 35 & 100.321 & 9.48475 & 18.83 & 2.15   & iv          \\
 36 & 100.278 & 9.59888 & 18.92 & 0.77   & -           \\
 37 & 100.256 & 9.57586 & 18.93 & 3.68   & iv          \\
 38 & 100.187 & 9.71719 & 18.99 & 1.72   & 2.31        \\
 39 & 100.516 & 9.81725 & 19.04 & 0.67   & -           \\
 40 & 100.319 & 9.50628 & 19.29 & 3.03   & -           \\
 41 & 100.339 & 9.90625 & 19.59 & 0.15   & 0.15        \\
 42 & 100.296 & 9.57014 & 19.78 & 3.03   & -           \\
 43 & 100.147 & 9.48655 & 19.88 & 0.16   & 0.16        \\
 44 & 100.170 & 9.74521 & 20.70 & 3.97   & -           \\ 
 \hline
 \end{tabular}
\end{table}

\begin{figure}
	\centering
  	\includegraphics[width=.45\textwidth]{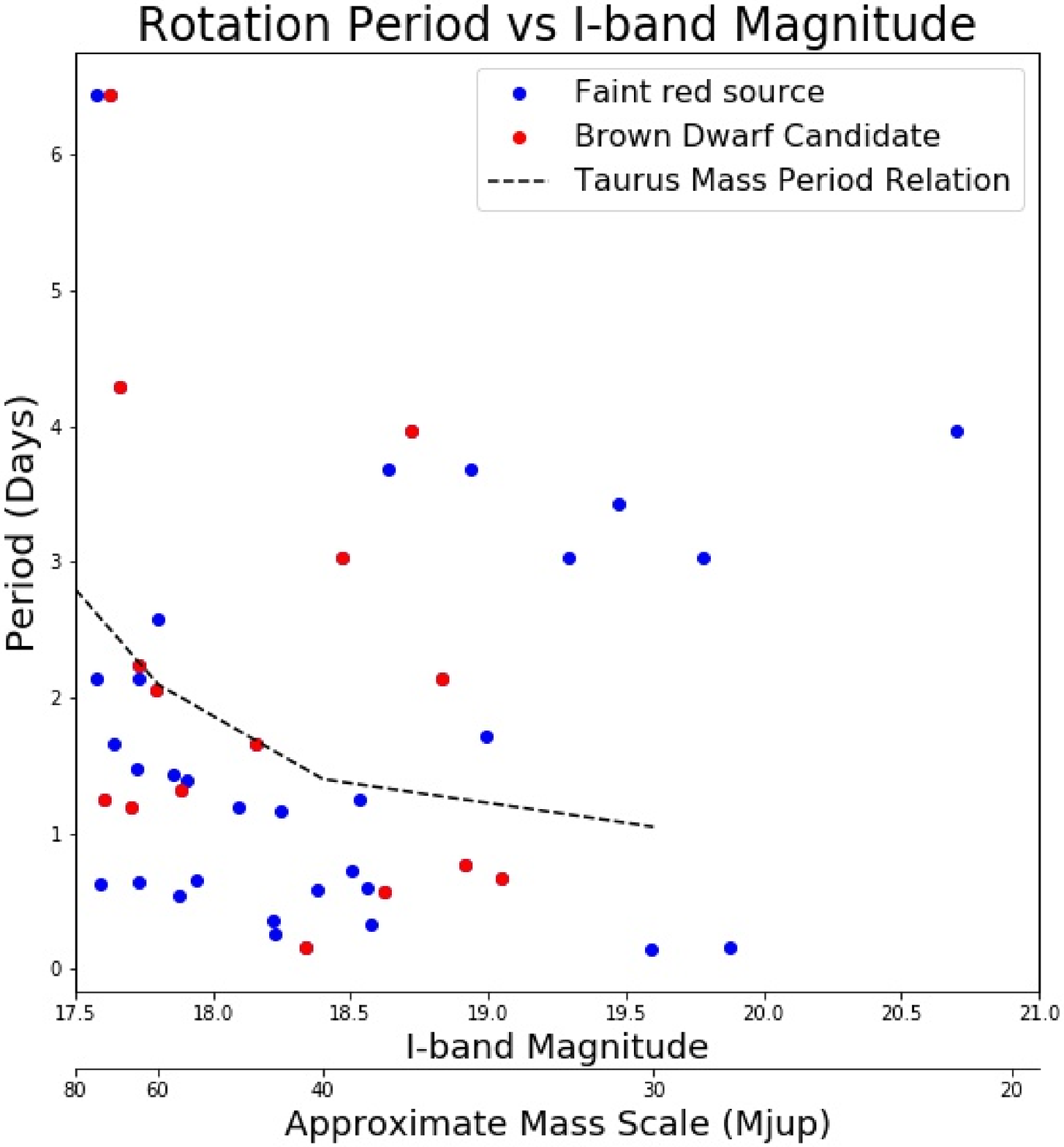}
	\caption{I-band vs. rotation period. The dashed black line indicates the period-mass relation found in Taurus by \citep{spinmass}}
	\label{fig:Rot2}
\end{figure} 

Figure \ref{fig:Rot2} shows the relation between I-band magnitude and period for our sample of 44 sources. A mass scale has been constructed using a 3Myr isochrone. The mass scale assumes zero extinction ($A_V = 0$) and as such, should only be viewed as an approximate indication. The dashed black line in Figure \ref{fig:Rot2} indicates the period-mass relation found in Taurus by \citet{spinmass}. Our sample does appear to roughly follow this trend of the median period decreasing with decreasing mass. It is difficult to extend this relation to very faint sources as for $I < 19$ ($\approx0.03M_\odot$) the sample size is very small. The large scatter in the periods even for the brighter sources is likely due to the young age as accretion and the initial formation conditions will still have a significant influence.

\subsection{Unusual variability}

\begin{figure*}
	\centering
    \includegraphics[width=1\textwidth]{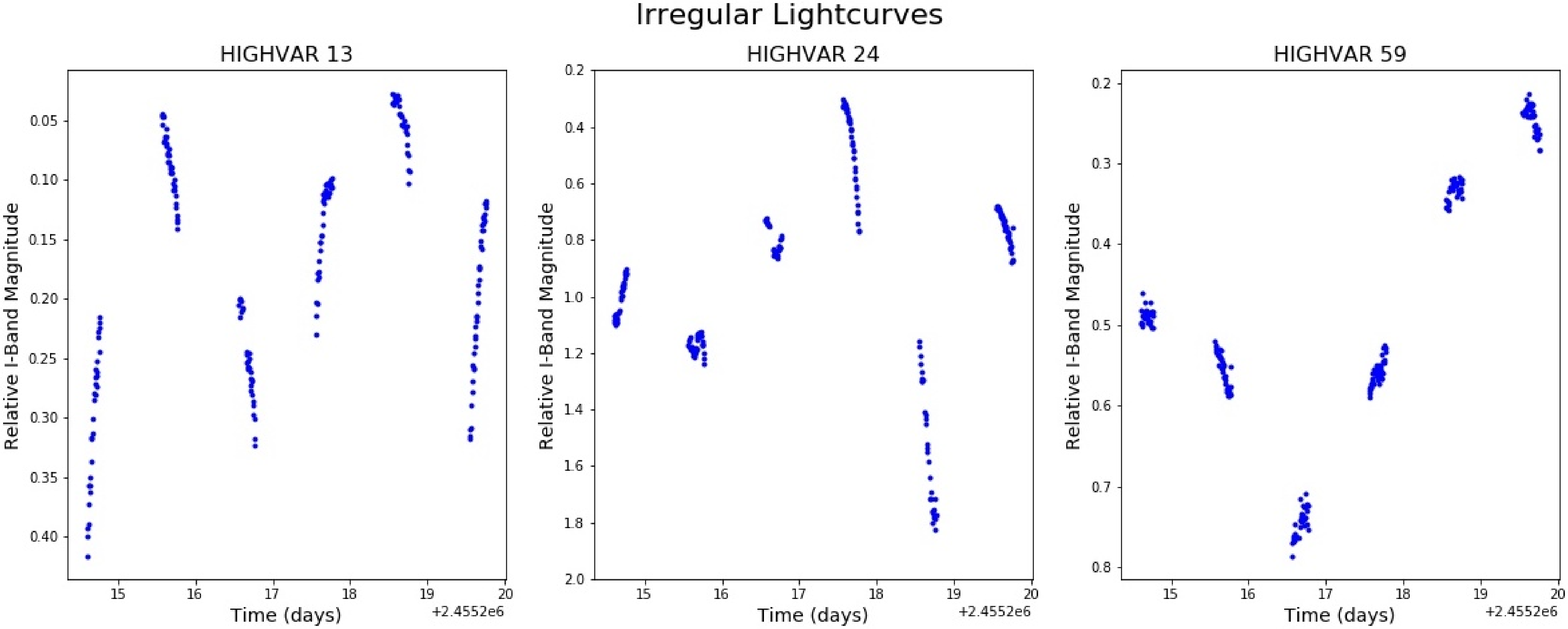}
	\caption{Lightcurves for three of the HIGHVAR sample, showing high amplitude and irregular variability over the six nights of observation.}
	\label{fig:Irreg}
\end{figure*}

Some of the sources from DISCS and HIGHVAR show particularly irregular lightcurves with high amplitude variation. In total, we find 22 objects with irregular amplitude variations greater than 0.2 magnitudes. A small sample of these lightcurves are shown in Figure \ref{fig:Irreg}. A high portion of this subsample (64\%) have mid-infrared excess and thus evidence of discs. For the remainder, a disc may still be present, but remains undetected at 3-5$\,\mu m$. We attribute this type of variability to accretion from the disc, caused by variable accretion rates, unstable hot spots, or obscurations by disc material, or a combination thereof. This type of variability is commonly observed in young stellar objects \citep{cody17}, and has also been found in the substellar domain \citep{aleksVar}. Indicating that magnetospheric accretion from a disc operates in brown dwarfs in a similar manner to stars.


\section{Conclusions}

We have used optical, near and mid-infrared photometry, time series data, extinction maps and to some extent Gaia kinematics to construct a catalogue of 902 faint red sources with indicators of youth for NGC2264. Within this catalogue we select 429 brown dwarf candidates based on their infrared colours. The brown dwarf candidates are estimated to span a mass range from 0.08 to 0.01\,$M_{\odot}$. This extends the brown dwarf candidate population of NGC2264 to much lower masses than previous studies have explored for this cluster ($\sim 0.04\,M_{\odot}$ \citet{sung2009}). Our sample has deep near-infrared photometry and colours for all objects, as well as time series data spanning six nights. We have determined rotation periods for 44 sources and classified the disc morphology of 199 sources. Having a large, yet well characterised sample of young brown dwarfs is essential for making progress on issues such as formation mechanisms and the shape of the very low mass IMF. This large sample of brown dwarf candidates is the first step towards these goals and has potential to produce one of the largest single samples of brown dwarfs in a cluster, once confirmed by follow up spectroscopy.



\section*{Acknowledgements}

We would like to thank Jerome Bouvier and Jonathan Irwin for their help when preparing the Blanco observations. We acknowledge support from STFC through grant number ST/R000824/1. K.M. acknowledges funding by the Science and Technology Foundation of Portugal (FCT), grants No. IF/00194/2015, PTDC/FIS-AST/28731/2017 and UIDB/00099/2020. This work is based [in part] on observations made with the Spitzer Space Telescope, which was operated by the Jet Propulsion Laboratory, California Institute of Technology under a contract with NASA. This work has made use of data from the European Space Agency (ESA) mission
{\it Gaia} (\url{https://www.cosmos.esa.int/gaia}), processed by the {\it Gaia}
Data Processing and Analysis Consortium (DPAC,
\url{https://www.cosmos.esa.int/web/gaia/dpac/consortium}). Funding for the DPAC
has been provided by national institutions, in particular the institutions
participating in the {\it Gaia} Multilateral Agreement.

\section*{Data Availability}
The first 5 rows of the faint red sources \& brown dwarf candidate table are shown in Table A1. The full table is available in the online supplementary material.




\bibliographystyle{mnras}
\bibliography{Bib.bib}




\appendix

\section{Brown Dwarf Candidates}

\begin{table*}
 \caption{The first 5 rows of the faint red sources table, brown dwarf candidates are indicated by an `X' in the BDC column. The full table is available online. An $X$ indicates that a source is found in the corresponding sample.}
 \label{tab:prob}
\begin{tabular}{lllllllllllllll}
 \hline
 RA         & DEC      & I     & J     & H     & K      & BDC & JHK\_Sel & IKJH\_Sel & IJ\_Sel & VAR & HIGHVAR & DISCS & KINEMATIC & HIGHEX \\
  \hline
 99.961919  & 9.560050 & 19.01 & 16.68 & 15.59 & 15.12  & -   & -        & -         & -       & -   & -       & X     & -         & -      \\
 99.964158  & 9.786519 & 20.90 & 18.20 & 17.22 & 16.62  & X   & X        & -         & X       & X   & X       & -     & -         & -      \\
 99.965286  & 9.653558 & 20.38 & 17.90 & 16.91 & 16.39  & X   & -        & -         & X       & X   & -       & -     & -         & -      \\
 99.965698  & 9.484825 & 17.87 & 16.12 & 88.89 & 100.00 & -   & -        & -         & -       & X   & -       & -     & -         & -      \\
 99.965967  & 9.776077 & 17.79 & 15.85 & 15.24 & 14.88  & X   & X        & -         & -       & X   & -       & -     & -         & -      \\
  \hline
 \end{tabular}
\end{table*}



\bsp	
\label{lastpage}
\end{document}